\documentclass[%
 reprint,%
 amssymb, amsmath,%
 aip,jcp,%
]{revtex4-1}

\usepackage{docs}%
\usepackage{bm}%
\usepackage[colorlinks=true,linkcolor=blue]{hyperref}%
\expandafter\ifx\csname package@font\endcsname\relax\else
 \expandafter\expandafter
 \expandafter\usepackage
 \expandafter\expandafter
 \expandafter{\csname package@font\endcsname}%
\fi
\usepackage{graphicx}
\begin{document}

\author{Johannes Fiedler}
\email{johannes.fiedler@physik.uni-freiburg.de}
\affiliation{Institute of Physics, Albert-Ludwigs University of Freiburg, Hermann-Herder-Str. 3, D-79104 Freiburg, Germany}
\affiliation{Centre for Materials Science and Nanotechnology,
Department of Physics, University of Oslo, P. O. Box 1048 Blindern,
NO-0316 Oslo, Norway}
\author{Clas Persson}
\affiliation{Centre for Materials Science and Nanotechnology,
Department of Physics, University of Oslo, P. O. Box 1048 Blindern,
NO-0316 Oslo, Norway}
\author{Mathias  Bostr{\"o}m}
\affiliation{Department of Energy and Process Engineering, Norwegian University of Science and Technology, NO-7491 Trondheim, Norway}

\author{Stefan Y. Buhmann}
\affiliation{Institute of Physics, Albert-Ludwigs University of Freiburg, Hermann-Herder-Str. 3, D-79104 Freiburg, Germany}
\affiliation{Freiburg Institute for Advanced Studies, Albert-Ludwigs-Universit{\"a}t Freiburg, Albertstr. 19, D-79104 Freiburg, Germany}
\title[Orientational dependence of the van der Waals interactions for finite-sized particles]
  {Orientational dependence of the van der Waals interactions for finite-sized particles}

\date{March 2010}%
\revised{August 2010}%
 \begin{abstract}
  Van der Waals forces as interactions between neutral and polarisable particles act at small distances between two objects. Their theoretical origin lies in the electromagnetic interaction between induced  dipole moments caused by the vacuum fluctuations of the ground-state electromagnetic field. The resulting theory well describes the experimental situation in the limit of the point dipole assumption. At smaller distances, where the finite size of the particles has to be taken into account, this description fails and has to be corrected by higher orders of the multipole expansion, such as quadrupole moments and so on. With respect to the complexity of the spatial properties of the particles this task requires a considerable effort. In order to describe the van der Waals interaction between such particles, we apply the established method of a spatially spread out polarisability distribution to approximate the higher orders of the multipole expansion. We hence construct an effective theory for effects from anisotropy and finite size on the van der Waals potential.
 \end{abstract}
\maketitle


\section{Introduction}
The interaction between two neutral but polarisable particles when they are closely brought together is described by van der Waals forces. They are caused by ground-state fluctuations of the electromagnetic field\cite{Buhmann2012} which result in a mostly attractive force between both particles. Typically, the interacting particles are considered to be point-like. However, current investigations and applications deal with large molecules and clusters on small separations, such as fullerenes (C$_{60}$ and C$_{70}$ with their derivatives) that find application in a plethora of different fields like nanomedicine\cite{doi:10.2217/nnm.13.99}, hydrogen storage\cite{doi:10.1021/nl071436g}, bio-organics\cite{Cheng2015} and photodynamic therapy\cite{B711141J} but particularly as n-type semiconductors in several branches of organic electronics\cite{brabec2011organic,Das17,laquai2015charge,kastner2013improvement}. In such experiments and applications the separation of the considered particles is of the same order of magnitude as their spatial dimension. Hence, the usual theory results in a incorrect estimates due to the breakdown of the point-particle assumption. In order to describe their results correctly an extension of the theory with respect to the finite size of the particle is required. This goes hand in hand with the extension of theory to account for orientation. Therefore, we start with an analysis of the orientational dependency of the van der Waals potential and introduce an effective eccentricity of the particle's polarisability which decreases the numerical effort on estimations of such forces. Based on its knowledge we discuss its influence on small anisotropic particles. Thereafter, we attend to van der Waals forces on extended and orientated particles and derive the resulting impact on the van der Waals potential. Here, we use an established model defining the size of a particle based on its electronic density\cite{PN09} which was successfully tested on ionic liquids\cite{doi:10.1021/jp212154c} and Casimir--Polder forces\cite{Fiedler15,Brand15}. In contrast to previous works, where the influence of extension or orientation was studied for atoms \cite{PSSB:PSSB2220860245,0305-4470-8-11-019}, specific molecules \cite{Priya15}, or in a nonperturbative way resulting in higher order polarisabilities \cite{Mtichell72,1402-4896-90-3-035405,PhysRevA.90.054502}, the introduced method yields a model which can be applied to numerous situations based on the dipole polarisabilities of the particles and the corresponding extension parameters.

\section{Particle's orientation}
First, we analyse the dependence of the van der Waals interaction on the orientation of the particle. To this end, we start with the van der Waals potential, that follows from the application of perturbation theory to the macroscopic electric dipole-electric field interaction $\hat{\bf{d}}\cdot {\hat{\bf{E}}}$ with respect to the ground-state of the electromagnetic field \cite{Buhmann2012}
\begin{eqnarray}
 \lefteqn{U_{vdW}({\bf{r}}_A,{\bf{r}}_B) = -\frac{\hbar\mu_0^2}{2\pi} \int\limits_0^\infty \mathrm d \xi \, \xi^4  \operatorname{tr} \left[ \boldsymbol{\alpha}_A (i\xi)\right.}\nonumber\\
 &&\left.\cdot {\bf{G}}({\bf{r}}_A,{\bf{r}}_B,i\xi) \cdot \boldsymbol{\alpha}_B (i\xi)\cdot {\bf{G}}({\bf{r}}_B,{\bf{r}}_A,i\xi)\right] \, ,\label{eq:Uvdw}
\end{eqnarray}
where we restrict ourselves to the electric part of the interaction. The integrand on the right hand side of the equation has to be read from right to left and means that a virtual photon with the frequency $i\xi$ is created at particle $A$, which is located at ${\bf{r}}_A$, and propagates towards the other particle $B$, located at ${\bf{r}}_B$. This propagation is described by the Green function, which is the fundamental solution of the vector Helmholtz equation for the electric field \cite{Buhmann2012}
\begin{equation}
 \nabla \times \nabla \times {\bf{E}}({\bf{r}})- \frac{\omega^2}{c^2} \varepsilon(\omega) {\bf{E}}({\bf{r}}) = i \mu_0 \omega {\bf{j}}({\bf{r}}) \, . 
\end{equation}
After the propagation, the virtual photon interacts with the particle $B$, which is expressed by the multiplication of its polarisability $\boldsymbol{\alpha}_B$. Afterwards, this photon is back scattered towards the first particle, which is again given by the Green tensor with exchanged coordinates. Finally, the back scattered photon interacts with the polarisability of the particle $A$, $\boldsymbol{\alpha}_A$. The sum over all these virtual photons results the van der Waals potential respectively the force. Usually, the polarisabilities will be averaged isotropically, resulting in
\begin{equation}
 {\boldsymbol{\alpha}}(\omega) =\alpha(\omega) {\mathbf{1}} \, ,
\end{equation}
with the unit matrix $\mathbf{1}$. This approximation is only valid for isotropic or weakly anisotropic molecules. In real situations the anisotropy has to be taken into account and influences the van der Waals interaction. In order to include the orientation of the particles in the van der Waals potential~(\ref{eq:Uvdw}), the different reference frames have to be considered. The dyadic Green function is given in the laboratory system, whereas the polarisabilities are in the molecular fixed frame. Hence these tensors have to be rotated with respect to the laboratory system:
\begin{equation}
 {\boldsymbol{\alpha}}_{A,B}(\omega, {\bf{\Omega}}) = {\bf{R}}({\bf{\Omega}}) \cdot {\boldsymbol{\alpha}}_{A,B}(\omega) \cdot {\bf{R}}^{-1}({\bf{\Omega}}) \, , \label{eq:rot}
\end{equation}
with the rotational matrix ${\bf{R}}$, that rotates the particle via the three rotational axes (the pitch, the yaw and the roll axis). In order to define the Euler angels for its description, it is useful to use Tait--Bryan angles\cite{Fiedler15,Landau}. In general, the polarisability is a fully occupied tensor. We assume that the off-diagonal elements are negligibly small
\begin{equation}
 \boldsymbol{\alpha} = \operatorname{diag}(\alpha_{xx},\alpha_{yy},\alpha_{zz}) \, ,
\end{equation}
they vanish in an appropriate molecular-fixed frame.

For the following considerations, we assume that the particles are embedded in a homogeneous media, described by the Green function for the bulk material \cite{Buhmann2012,PhysRevA.78.053828,Scheel2008}
\begin{eqnarray}
\lefteqn{ {\bf{G}}({\bf{r}},{\bf{r}}',\omega) = -\frac{1}{3k^2} \boldsymbol{\delta}(\boldsymbol{\varrho})- \frac{\mathrm e^{ik\varrho}}{4\pi k^2 \varrho^3}}\nonumber\\
&&\times\left\lbrace \left[ 1-ik\varrho-(k\varrho)^2\right]{\mathbf{1}}-\left[3-3ik\varrho -(k\varrho)^2\right]{\bf{e}}_\varrho \otimes {\bf{e}}_\varrho\right\rbrace \, ,
\end{eqnarray}
with wave number $k=\sqrt{\varepsilon(\omega)}\omega/c$, relative coordinate $\boldsymbol{\varrho} = {\bf{r}}-{\bf{r}}'$ and its absolute value and its unit vector, $\varrho=\left|\boldsymbol{\varrho}\right|$ and ${\bf{e}}_\varrho = \boldsymbol{\varrho}/\varrho$, respectively. The bold $\delta$-function denotes the product of the Dirac-$\delta$-function and the unit tensor ${\bf{I}}$, $\boldsymbol{\delta}(\boldsymbol{\varrho})={\delta}(\boldsymbol{\varrho}){\bf{I}}$. For short distances, this expression can be expanded in series ($\omega\varrho/c\ll1$) and results the nonretarded Greens function
\begin{equation}
 {\bf{G}}(\boldsymbol{\varrho},\omega) \approx -\frac{1}{3k^2}\boldsymbol{\delta}(\boldsymbol{\varrho}) -\frac{1}{4\pi k^2\varrho^3}\left(\mathbf{1}-3{\bf{e}}_\varrho\otimes{\bf{e}}_\varrho\right) \, .
\end{equation}
For simplicity, we chose one particle is located at the origin ${\bf{r}}=(0,0,0)^T$ and the other particle along the $x$-axis, ${\bf{r}}'=(x,0,0)^T$, resulting in a dyadic Green function
\begin{equation}
 {\bf{G}}(x,\omega) = -\frac{1}{4\pi k^2 x^3} \operatorname{diag} (-2,1,1) \, . \label{eq:Gscat}
\end{equation}

By considering a point dipole and its orientations, the roll angle has not to be considered. In the following, we analyse the dependence of the van der Waals potential with respect to a rotation around the yaw and the pitch axis separately.

\subsection{The molecular polarisability eccentricity}
\begin{figure}[htb]
 \includegraphics[width=\columnwidth]{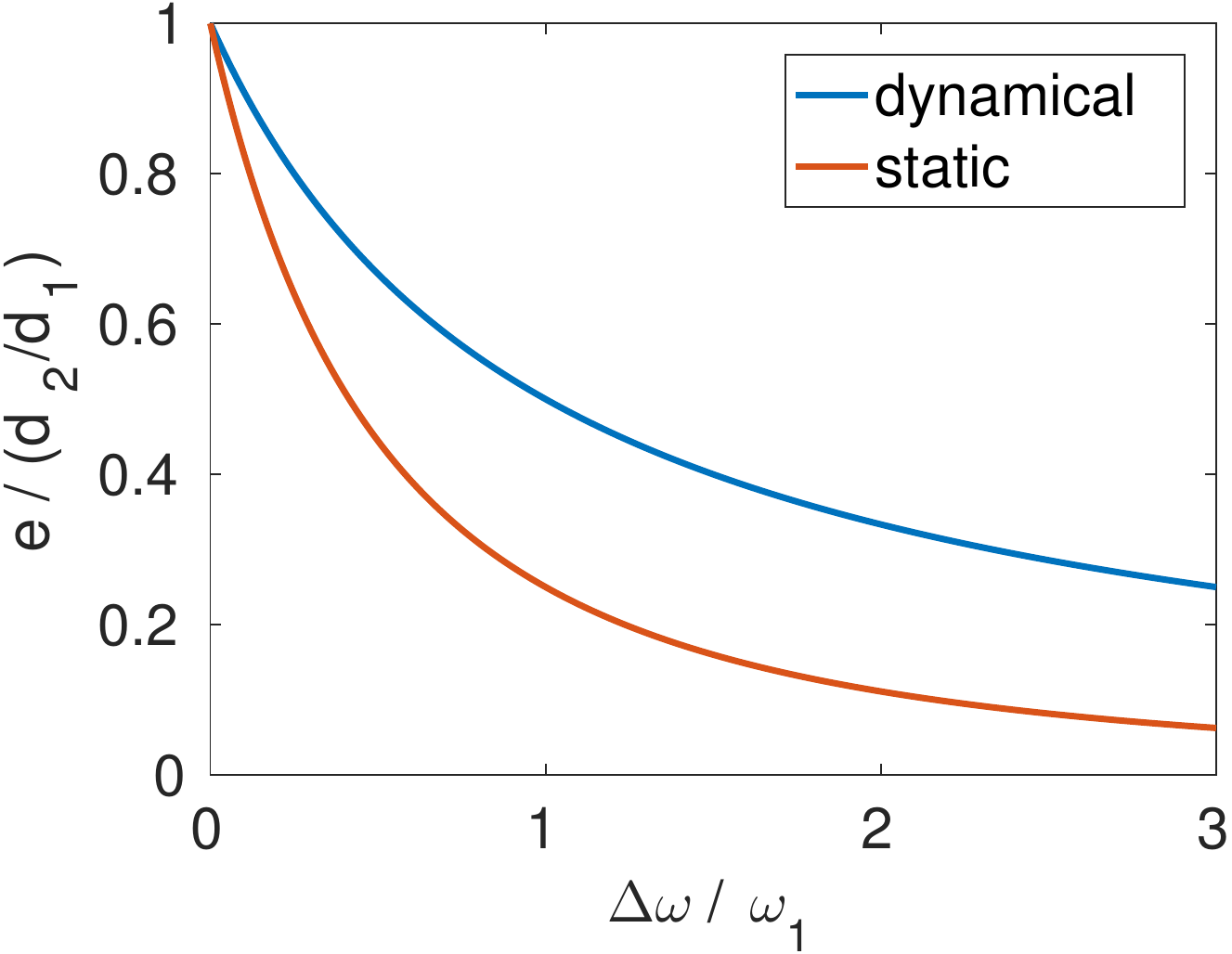}
 \caption{Comparison of the different eccentricities defined by the integral method (blue line - dynamical) and the static method (red line - static).} \label{fig:eccentricity}
\end{figure}

In principle, the elements of a molecular polarisability tensor are independent from each other, because of the coupling to different directions of the induced transition dipole moment. Hence, they have to be taken separately in order to describe the van der Waals interaction of an anisotropic particle. In order to estimate the general behaviour of this interaction it is useful to define the eccentricities. In general, all elements differ from each other, necessitating the introduction of three eccentricities $e_1$, $e_2$ and $e_3$ in such way that the polarisability reads
\begin{equation}
 \boldsymbol{\alpha} = \operatorname{diag}(\alpha_{xx},\alpha_{yy},\alpha_{zz}) = \alpha_0(\omega) \operatorname{diag}(e_1,e_2,e_3) \, ,
\end{equation}
with a purely frequency dependent part $\alpha_0(\omega)$ and a constant diagonal matrix with the eccentricities. 
Different ways exist for the determination of these parameters. One is to take one element of the polarisability tensor as the frequency dependent part, e.g. $\alpha_0(\omega) = \alpha_{xx}(\omega)$; this results in the eccentricities $e_1=1$, $e_2 = \alpha_{yy}/\alpha_{xx}$ and $e_3=\alpha_{zz}/\alpha_{xx}$. Another possibility is the use of the scalar polarisability (the trace over the tensor) as the frequency dependent part, that leads to $\alpha_0 = [\alpha_{xx}+\alpha_{yy}+\alpha_{zz}]/3$ and the corresponding eccentricities $e_i = \alpha_{ii}/\alpha_0$. In general, as it is indicated above, the eccentricities still depend on the frequency. One often uses static polarisabilities to define them. With respect to the estimation of dispersion forces on the other hand, this may yield a large deviation from the exact forces evaluated with the complete frequency dependency. In order to obtain this deviation, one has to define the dynamical eccentricities suitable for the van der Waals potential:
\begin{equation}
 e_i = \frac{\int_0^\infty \, \alpha_{ii}(\omega)}{\int_0^\infty \, \alpha_{0}(\omega)} \, .
\end{equation}
In order to illustrate the difference between this eccentricity and the one evaluated with the static values, we assume a single resonance in each polarisability at $\omega_1$ and $\omega_2$ with the associated oscillator strengths $d_1$ and $d_2$, which results in the polarisabilities
\begin{equation}
 \alpha_{ii}(i\xi) = d_{i}(\xi^2 + \omega_i^2)^{-1} \, ,
\end{equation}
for $i=\{x,z\}$, which is a Drude model. Here, we have rotated the frequency integral from the real to the imaginary axis as  is used for the evaluation of the dispersion forces. Using this model the eccentricity reads
\begin{equation}
 e=\frac{d_2}{d_1} \frac{\omega_1}{\omega_1 + \Delta\omega}\, ,
\end{equation}
with the difference frequency $\Delta\omega=\omega_2-\omega_1$. Using the static polarisability instead the eccentricity is given by
\begin{equation}
 e_0 = \frac{d_2}{d_1}\frac{\omega_1^2}{(\omega_1+\Delta\omega)^2}\, .
\end{equation}
The difference between both can be seen in Fig.~\ref{fig:eccentricity}. For a small detuning between both resonances ($\Delta\omega\to0$) both results are getting close together and equal the ratio between the corresponding oscillator strengths $d_2/d_1$. For large separations, the correct dynamical approach results in approximately twice the static result.

\begin{figure}[htb]
 \includegraphics[width=0.8\columnwidth]{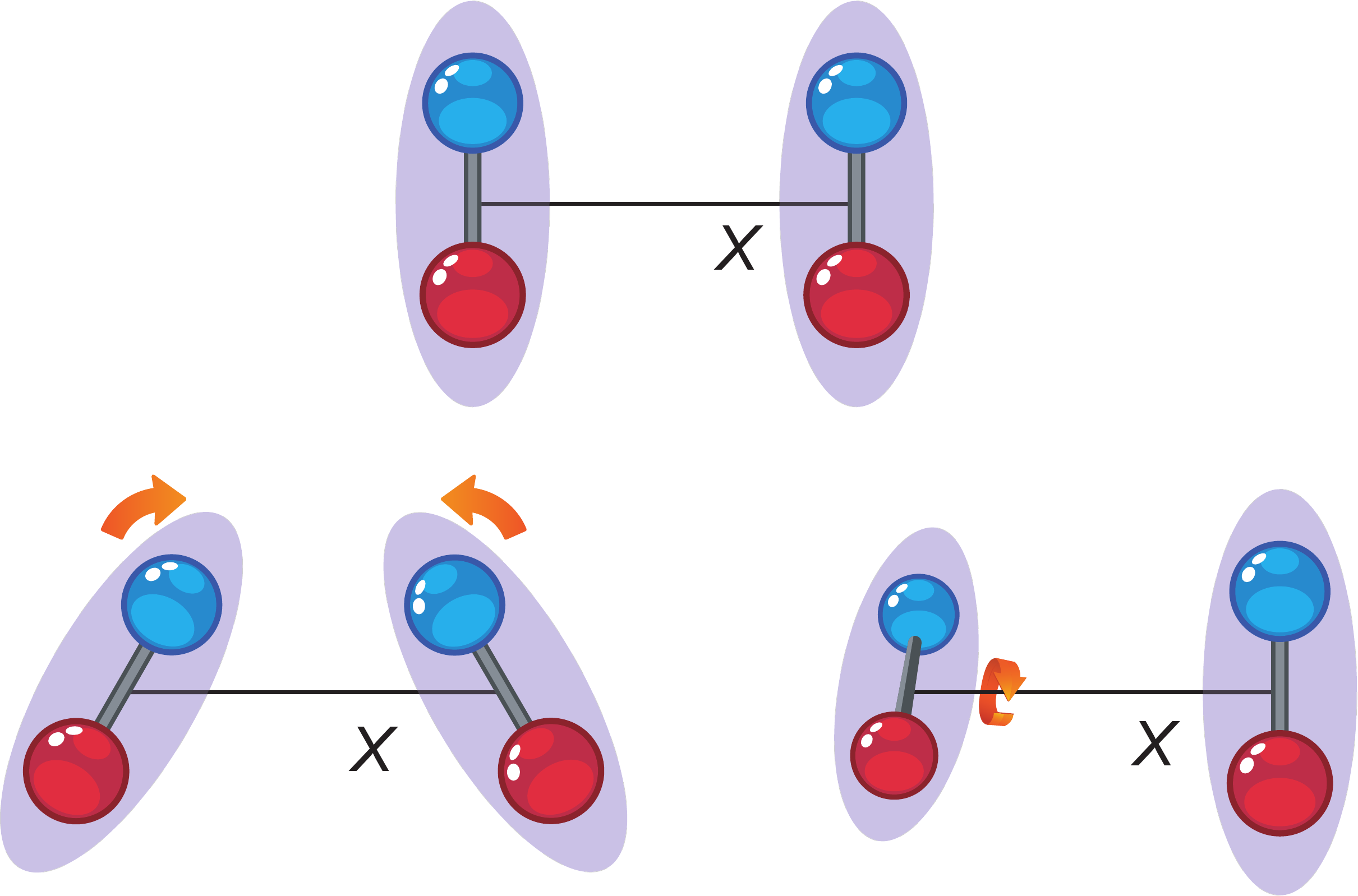}
 \caption{Sketch of the considered orientations. The original alignment is depicted in the top picture. The rotation inside the plane of the molecules (rotation around the yaw axis) is shown on the bottom left. The rotation around the axis connecting the centres of both particles (rotation around the pitch axis) is shown on the bottom right.}\label{fig:rotations}
\end{figure}

\subsection{Rotation around the yaw axis}
A rotation around the yaw axis is equivalent to a shearing of the system (see Fig.~\ref{fig:rotations} bottom left) and is a rotation around the $y$-axis by the angle $\vartheta$ with the rotation matrix
\begin{equation}
 {\bf{R}}(\vartheta) = \begin{pmatrix}
                        \cos\vartheta & 0 & \sin\vartheta \\
                        0 & 1 & 0 \\
                        -\sin\vartheta & 0 & \cos\vartheta
                       \end{pmatrix} \, .
\end{equation}
Combining Eqs.~(\ref{eq:Gscat}), (\ref{eq:rot}) with the rotation matrix and inserting them into the van der Waals potential, Eq.~(\ref{eq:Uvdw}) results
\begin{eqnarray}
\lefteqn{ U_{vdW}(x,\vartheta,\vartheta') = -\frac{\hbar\mu_0^2}{32\pi^3x^6} \int\limits_0^\infty \mathrm d \xi \, \frac{\xi^4}{k^4} 
 \operatorname{tr} \left[ {\bf{R}}(\vartheta)\right.}\nonumber \\&&\left.\cdot \boldsymbol{\alpha}_A (i\xi)\cdot{\bf{R}}(-\vartheta)\cdot \operatorname{diag} (-2,1,1)
\cdot{\bf{R}}(\vartheta')\cdot \boldsymbol{\alpha}_B (i\xi)\right.\nonumber \\&&\left.\cdot {\bf{R}}(-\vartheta')\cdot \operatorname{diag} (-2,1,1)\right] \, , \label{eq:Uvdw2}
\end{eqnarray}
where $\vartheta$ and $\vartheta'$ denotes the rotation angles of the particle $A$ and $B$, respectively. For simplicity, we assume a symmetric top with the polarisability
\begin{equation}
 \boldsymbol\alpha_{A,B}(i\xi) =\alpha(i\xi)_{A,B} \operatorname{diag}(e,e,1) \, , 
\end{equation}
with the eccentricity $e$ that defines an ellipsoidal molecule that is prolate for $e<1$ and oblate for $e>1$ and is equal for both particles. Due to the product ansatz of the polarisability the frequency dependence factorises and the orientational dependency can be written as a correction of the van der Waals potential
\begin{eqnarray}
\lefteqn{ U_{vdW}(x,\vartheta,\vartheta') = U_{vdW}(x) \left[1  +\frac{5}{4}\cos^2\vartheta\cos^2\vartheta'\right.}\nonumber \\ &&\left.-\cos\vartheta\sin\vartheta\cos\vartheta'\sin\vartheta'-\cos^2\vartheta-\cos^2\vartheta'\right.\nonumber \\
&&\left.+\left(-\frac{5}{2}\cos^2\vartheta\cos^2\vartheta'+2\cos\vartheta\sin\vartheta\cos\vartheta'\sin\vartheta'\right.\right.\nonumber\\
&&\left.\left.+\frac{5}{4}\cos^2\vartheta+\frac{5}{4}\cos^2\vartheta'\right)e + \left(\frac{5}{4}\cos^2\vartheta\cos^2\vartheta'\right.\right.\nonumber\\
&&\left.\left.-\cos\vartheta\sin\vartheta\cos\vartheta'\sin\vartheta'+\frac{1}{4}\sin^2\vartheta+\frac{1}{4}\sin^2\vartheta'\right)e^2\right] \, ,
\end{eqnarray}
with the van der Waals potential for isotropic particles
\begin{equation}
 U_{vdW}(x) = -\frac{\hbar}{16\pi^3 \varepsilon_0^2 x^6} \int\limits_0^\infty \mathrm d \xi \, \alpha_A(i\xi)\alpha_B(i\xi) \,.
\end{equation}

For counterrotating molecules $\vartheta =-\vartheta'$ this result simplifies to
\begin{eqnarray}
\lefteqn{ \frac{U_{vdW}(x,\vartheta,-\vartheta)}{ U_{vdW}(x)} =1}\nonumber\\
&&+ \left(\frac{1}{4}\cos^4\vartheta+\frac{1}{2}\cos^2\vartheta+\frac{1}{2}\right)(e^2-1)\nonumber\\
&&+
 \left(-\frac{1}{2}\cos^4\vartheta+\frac{1}{2}\cos^2\vartheta\right)(e-1)  \, .
\end{eqnarray}
\begin{figure}[htb]
 \includegraphics[width=\columnwidth]{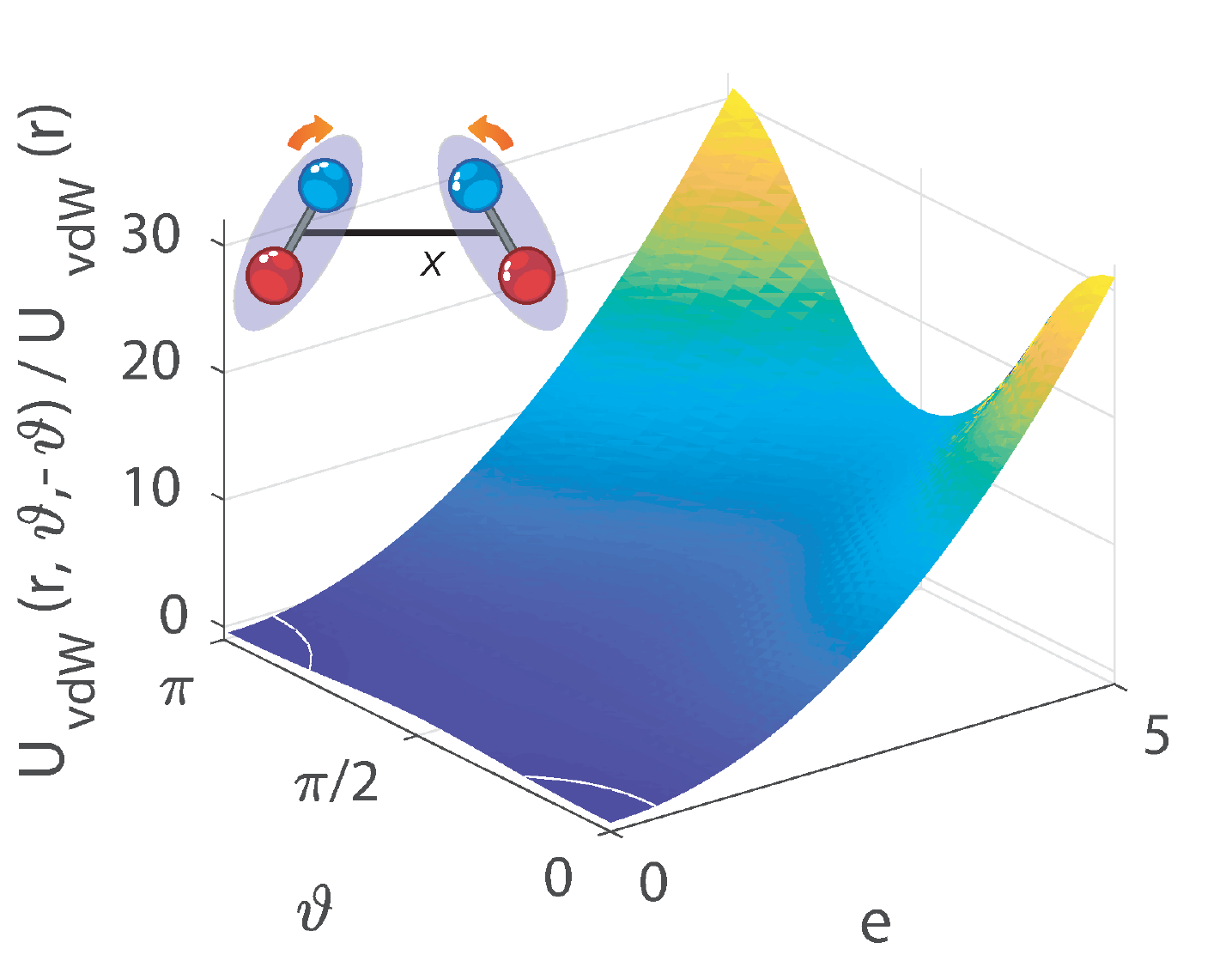}
 \caption{Sketch of the correction of the van der Waals interaction depending on the eccentricity $e$ and the counterrotating angle $\vartheta$. The white line denotes the vanishing of the van der Waals force.}\label{fig:counterrotating}
\end{figure}
Figure~\ref{fig:counterrotating} illustrates the correction of the van der Waals potential for the counterrotating case. It can be seen that the interaction rapidly increases with eccentricity, whereas the orientation results in the expected periodicity of $\pi$. 
A small region of parameters can be identified where the van der Waals force turns from attractive to repulsive. This is the case when the eccentricity assumes low values which means a prolate geometry of the considered particle and an angle between both particles close to $0^\circ$ or $90^\circ$. Hence, the prolate particles should be arranged almost parallel to achieve repulsion.
\subsection{Rotation around the pitch axis}\label{sec:pitch}
\begin{figure}[htb]
\includegraphics[width=\columnwidth]{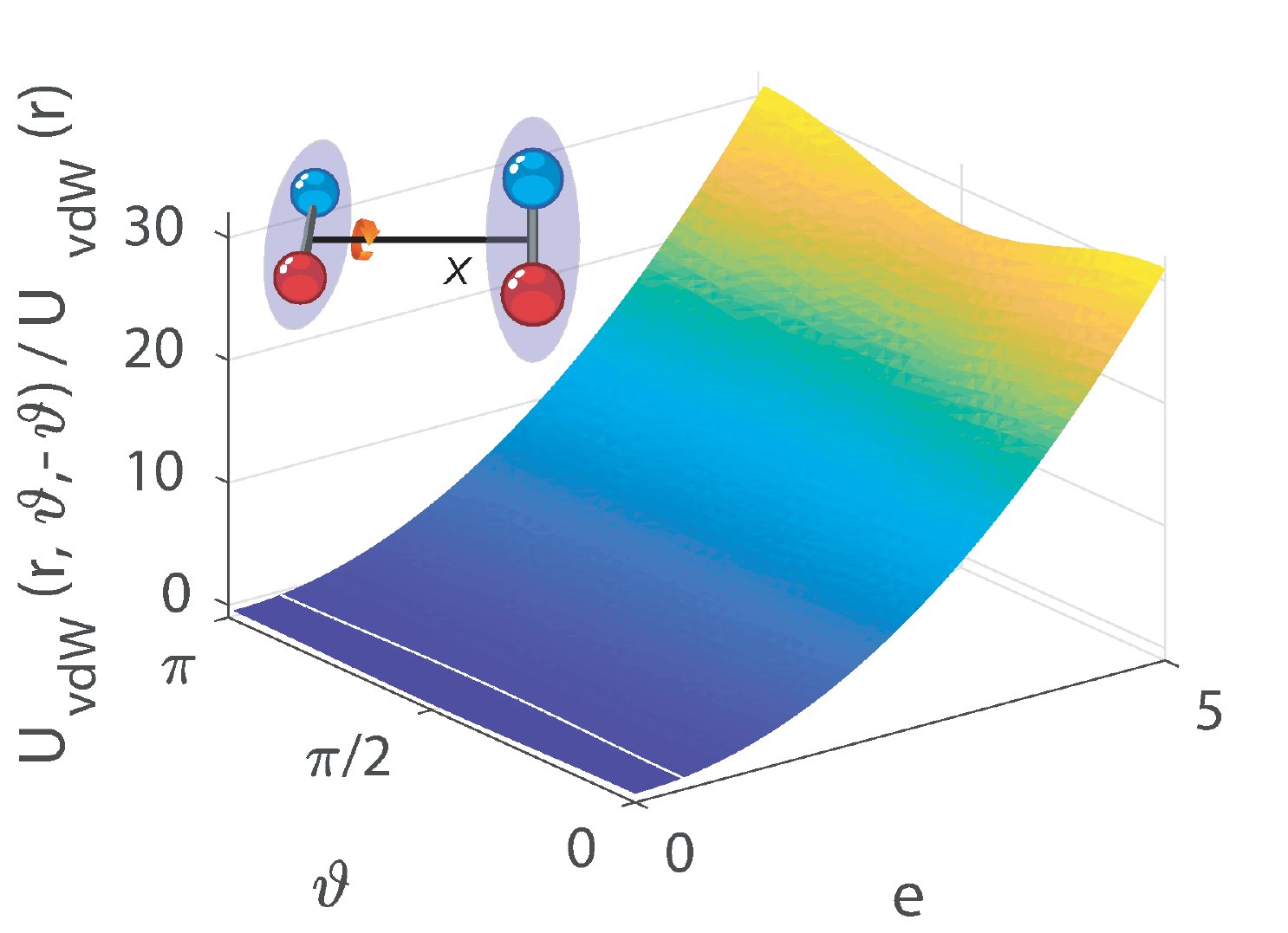}
\caption{Sketch of the correction of the van der Waals potential depending on the eccentricity $e$ and the torsion angle $\vartheta$. The white line denotes the vanishing of the van der Waals force.}\label{fig:torsion}
\end{figure}

A rotation around the pitch axis is equivalent to a torsion of the system (see Fig.~\ref{fig:rotations} bottom right) and is given by an elementary rotation around the $x$-axis with a rotation matrix
\begin{equation}
 {\bf{R}}(\vartheta) = \begin{pmatrix}
                        1 & 0 & 0 \\
                        0 & \cos\vartheta & -\sin\vartheta\\
                        0 & \sin\vartheta & \cos\vartheta
                       \end{pmatrix}
\, .
\end{equation}
This matrix has to be applied to the van der Waals potential, Eq.~(\ref{eq:Uvdw2}). Due to the symmetry of the system with respect to the $x$-axis it is sufficient to consider particle $A$ to be rotated and particle $B$ is fixed in the laboratory frame. This results in a van der Waals potential
\begin{eqnarray}
\lefteqn{ U_{vdW}(x,\vartheta) = -\frac{\hbar\mu_0^2}{32\pi^3x^6} \int\limits_0^\infty \mathrm d \xi \, \frac{\xi^4}{k^4}\operatorname{tr} \left[ {\bf{R}}(\vartheta)\cdot \boldsymbol{\alpha}_A (i\xi)\right. }\nonumber \\
&& \left.\cdot{\bf{R}}(-\vartheta)\cdot \operatorname{diag} (-2,1,1)\cdot\boldsymbol{\alpha}_B (i\xi)\cdot \operatorname{diag} (-2,1,1)\right] \, , 
\end{eqnarray}
leading to
\begin{eqnarray}
 \lefteqn{\frac{U_{vdW}(x,\vartheta)}{U_{vdW}(x)}= 1  +\left(-\frac{1}{2}\cos^2\vartheta+\frac{1}{2}\right)(e-1)}\nonumber\\&&+ \left(\frac{1}{4}\cos^2\vartheta+1\right)(e^2-1) \, .\qquad
\end{eqnarray}
Figure~\ref{fig:torsion} illustrates this result. Comparing with the shearing case, one finds again an increasing of the potential with respect to an increasing eccentricity of the particles. The angular dependency is much less pronounced than in the shearing case. 

One can find again parameters which creates a repulsive van der Waals force for anisotropic ground-state particles. This is consistent with a repulsive Casimir--Polder force, the force acting between a neutral particle with a macroscopic dielectric body, which can also yield repulsion due to the anisotropy of the particles \cite{Fiedler15}.

\section{Van der Waals interaction between extended particles}
When describing the van der Waals interaction of two particles  at small length scales where the separation is of the same order of magnitude as the extension of the particles, the point dipole representation of the particles leads to large errors in estimations. The standard way out to treat this problem is to take into account higher orders of the multipole expansion, e.g. quadrupoles and so on. While their estimation leads to an enormous numerical effort and are virtually impossible for huge particles, such as large organic molecules or clusters, a simpler approximation of these extension effects is the application of a spatially slurred out polarisability, as it is established in the theory for ionic liquids \cite{PN09} and Casimir--Polder interactions \cite{Fiedler15,Brand15}. In these publications, the authors introduce a Gaussian distributed spatial polarisability density
\begin{equation}
 \alpha(\varrho) = \frac{1}{\mathcal{N}}\mathrm{exp}\left[-\left(\frac{x^2}{a_x^2}+\frac{y^2}{a_y^2}+\frac{z^2}{a_z^2}\right)\right] \, ,
\end{equation}
with the normalisation 
\begin{eqnarray}
\lefteqn{\mathcal N= \int \mathrm d^3 r \exp\left\{ -\left( \frac{x^2}{a_x^2} +
 \frac{y^2}{a_y^2}+ \frac{z^2}{a_z^2}\right) \right\}}\nonumber\\
 &&=\pi^{3/2} a_xa_ya_z \, . \qquad \qquad\qquad
\end{eqnarray}
The polarisability is
assumed to be Gaussian distributed over an ellipsoidal volume $V$ with the main
axes $a_x$, $a_y$ and $a_z$, and whose surface matches the equipotential surface of
the electron density set to a low threshold value. An example determining these main axes in given in Ref.~\cite{Fiedler15}.
The dye molecule Phthalocyanine was considered and one finds the main axes $a_x=a_y=17 a_B$ and $a_z=3.5 a_B$. 


With such a spatially distributed polarisability density $\alpha(\varrho)$, one can introduce the complete polarisability of the particle by its multiplication with the frequency dependent part
\begin{equation}
 \boldsymbol{\alpha}(\varrho,\omega) = \alpha(\varrho) \cdot \boldsymbol{\alpha}(\omega) \, .\label{eq:prod}
\end{equation}
With respect to the Born series expansion of the scattering Green function \cite{Scheel2008,Buhmann2012} the resulting van der Waals potential has to be integrated over the particle volumes
\begin{eqnarray}
\lefteqn{ U_{vdW}({\bf{r}},{\boldsymbol{\Omega}}_1,{\boldsymbol{\Omega}}_2)}\nonumber\\
&&= \int\limits_{V_1} \mathrm d^3 \varrho' \int\limits_{V_2} \mathrm d^3 \varrho'' U_{vdW}({\bf{r}},{\boldsymbol{\Omega}}_1,{\boldsymbol{\Omega}}_2,{\boldsymbol{\varrho}}',{\boldsymbol{\varrho}}'') \, , \label{eq:UvdWVol}
\end{eqnarray}
with the point wise van der Waals potential
\begin{eqnarray}
\lefteqn{ U_{vdW}({\bf{r}},{\boldsymbol{\Omega}}_1,{\boldsymbol{\Omega}}_2,{\boldsymbol{\varrho}}',{\boldsymbol{\varrho}}'') = -\frac{\hbar\mu_0^2}{32\pi^3} \int\limits_0^\infty \mathrm d \xi \, \frac{\xi^4}{k^4} }\nonumber \\
&&\times \operatorname{tr} \left[ {\bf{R}}({\boldsymbol{\Omega}}_1)\cdot \boldsymbol{\alpha}_A ({\boldsymbol{\varrho}}',i\xi)\cdot{\bf{R}}^T({\boldsymbol{\Omega}}_1)\cdot \operatorname{diag} (-2,1,1)\right.\nonumber \\
&&\left.\cdot{\bf{R}}({\boldsymbol{\Omega}}_2)\cdot \boldsymbol{\alpha}_B ({\boldsymbol{\varrho}}'',i\xi)\cdot {\bf{R}}^T({\boldsymbol{\Omega}}_2)\cdot \operatorname{diag} (-2,1,1)\right] \nonumber \\
&&\times \frac{1}{\left|{\bf{r}} -{\bf{R}}^T({\boldsymbol{\Omega}}_1){\boldsymbol{\varrho}}' -{\bf{R}}^T({\boldsymbol{\Omega}}_2){\boldsymbol{\varrho}}''  \right|^6} \, ,
\end{eqnarray}
where we have used the nonretarded free-space scattering Green function, because the considered effects ensure on small length scales. Note that due to the scattering Green function the relative coordinate between both particles is along the $x$-direction, ${\bf{r}}=(x,0,0)$. The polarisabilities are given in the molecular fixed frame and by the multiplication with the rotation matrices from left and right they transform into the laboratory fixed frame. In general the rotations depend on three different Euler angles, that are written in a vector form ${\boldsymbol{\Omega}}_i$ for shortens. By using the product ansatz for the polarisabilities to separate between the frequency and the spatial dependencies and by applying the Taylor series expansion to the, which means that the distance between the centres of the particles $\bf{r}$ is larger than the particles radii $\boldsymbol{\varrho}'$ and $\boldsymbol{\varrho}''$
\begin{eqnarray}
\lefteqn{ \frac{1}{\left|{\bf{r}} -{\bf{R}}^T({\boldsymbol{\Omega}}_1){\boldsymbol{\varrho}}' -{\bf{R}}^T({\boldsymbol{\Omega}}_2){\boldsymbol{\varrho}}''  \right|^6} =}\nonumber \\&& \frac{1}{r^6} \frac{1}{\left|{\bf{e}}_r-\frac{\varrho'}{r}{\bf{R}}^T({\boldsymbol{\Omega}}_1){\bf{e}}_{\varrho'}-\frac{\varrho''}{r}{\bf{R}}^T({\boldsymbol{\Omega}}_2){\bf{e}}_{\varrho''}\right|^6} \,,
\end{eqnarray}
which can be series expanded for $\varrho',\varrho''\ll r$. The complete rotations are products of elementary rotations around particle's pitch axis ${\bf{R}}_x$, yaw axis ${\bf{R}}_y$ and roll axis ${\bf{R}}_z$ that results in \cite{Landau}
\begin{equation}
 {\bf{R}}({\boldsymbol{\Omega}}_i) = {\bf{R}}_x(\beta_1^i)\cdot {\bf{R}}_y(\beta_2^i)\cdot{\bf{R}}_z(\beta_3^i) \label{eq:rotprod}\, .
\end{equation}
Equivalent to the previous sections, the anisotropy of the particles can be written via the eccentricities
\begin{equation}
 {\boldsymbol{\alpha}}_{A,B}(\omega) =\alpha_{A,B}(\omega) \operatorname{diag}(e_1^{A,B},e_2^{A,B},1) \, , \label{eq:eccpol}
\end{equation}
that are defined with respect to the main axis along the $z$ direction, $\alpha_{zz}$. These eccentricities lead to use ellipsoidal coordinates describing the relative coordinates of the particles which yields the coordination transform\cite{Fiedler15}
\begin{eqnarray}
 x_{A,B}&=& \varrho_{1,2} e_1^{A,B}\sin\vartheta_{1,2}\cos\varphi_{1,2} \,,\\ 
  y_{A,B}&=& \varrho_{1,2}e_2^{A,B} \sin\vartheta_{1,2}\sin\varphi_{1,2}\,, \\
   z_{A,B}&=& \varrho_{1,2} \cos\vartheta_{1,2}\,,
\end{eqnarray}
which yields the volume element in this ellipsoidal coordinates $\mathrm d V_i = \varrho_i^2 e_1^ie_2^i\sin\vartheta_i\mathrm d \varrho_i\mathrm d \vartheta_i\mathrm d \varphi_i$ which are defined in analogy to spherical coordinates. Together with these coordinates, the product ansatz (\ref{eq:prod}) for the polarisabilities the eccentricities (\ref{eq:eccpol}), and the rotation matrices (\ref{eq:rotprod}), the volume averaged van der Waals potential (\ref{eq:UvdWVol}), can be determined and yields
\begin{eqnarray}
\lefteqn{\frac{ U_{vdW}({\bf{r}},{\boldsymbol{\Omega}}_1,{\boldsymbol{\Omega}}_2)}{U_{vdW}({\bf{r}})}=} \nonumber\\ &&g
\sum_{n,m} f_{nm}({\boldsymbol{\Omega}}_1,{\boldsymbol{\Omega}}_2,e_1^{A,B},e_2^{A,B})\frac{a_A^{2n}a_B^{2m}}{r^{2(n+m)}}\, , \label{eq:full}
\end{eqnarray}
where $a$ denotes the main axis along the $x$ direction.  

These functions $f_{nm}({\boldsymbol{\Omega}}_1,{\boldsymbol{\Omega}}_2,e_1^{A,B},e_2^{A,B})$ are algebraic equations depending on the ten parameters. It is remarkable that only even orders contribute which is consistent to previous results\cite{Fiedler15} and shows the correspondence to the multipole expansion of the electromagnetic field. Due to their lengthy construction we omit their explicit form, see App.~\ref{app1}, and discuss one special case of two identical particles with a symmetric polarisability tensor $e_1^A=e_2^A=e_1^B=e_2^B=e$ which counterwise rotate around their pitch axes ${\boldsymbol{\Omega}}_1 = (\vartheta,0,0)=-{\boldsymbol{\Omega}}_2$, this scenario is equivalent to the one from Sec.~II B. In this case one finds
\begin{eqnarray}
g&=& 1+\left(\frac{1}{4}\cos^4\vartheta+\frac{1}{2}\cos^2\vartheta+\frac{1}{2}\right)(e^2-1)\nonumber\\
&&+
 \left(-\frac{1}{2}\cos^4\vartheta+\frac{1}{2}\cos^2\vartheta\right)(e-1)\,,\\
 f_{0,0} &=& e^4\,,\\
 f_{1,0}&=&f_{0,1} = 12(e^2-1)\cos^2\vartheta-3e^2+21/2\,,\\
 f_{1,1} &=& 144(e^2-1)^2 \cos^4\vartheta-12 (e^2-1)(2e^2-27)\cos^2\vartheta\nonumber\\&& +24e^4 -108 e^2 + 189\,.
\end{eqnarray}
As a consistency check, one finds the ordinary van der Waals interaction by increasing the particle's extension $a_A,a_B\to0$ and setting the corresponding eccentricity to one $e=1$. 

The dye molecule Phthalocyanine has been studied in recent experiments~\cite{Brand15} and theories~\cite{Fiedler15}. In the latter reference one finds its eccentricity $e=0.2$ and a main axis of $a=17a_B$. 
\begin{figure}[htb]
 \includegraphics[width=\columnwidth]{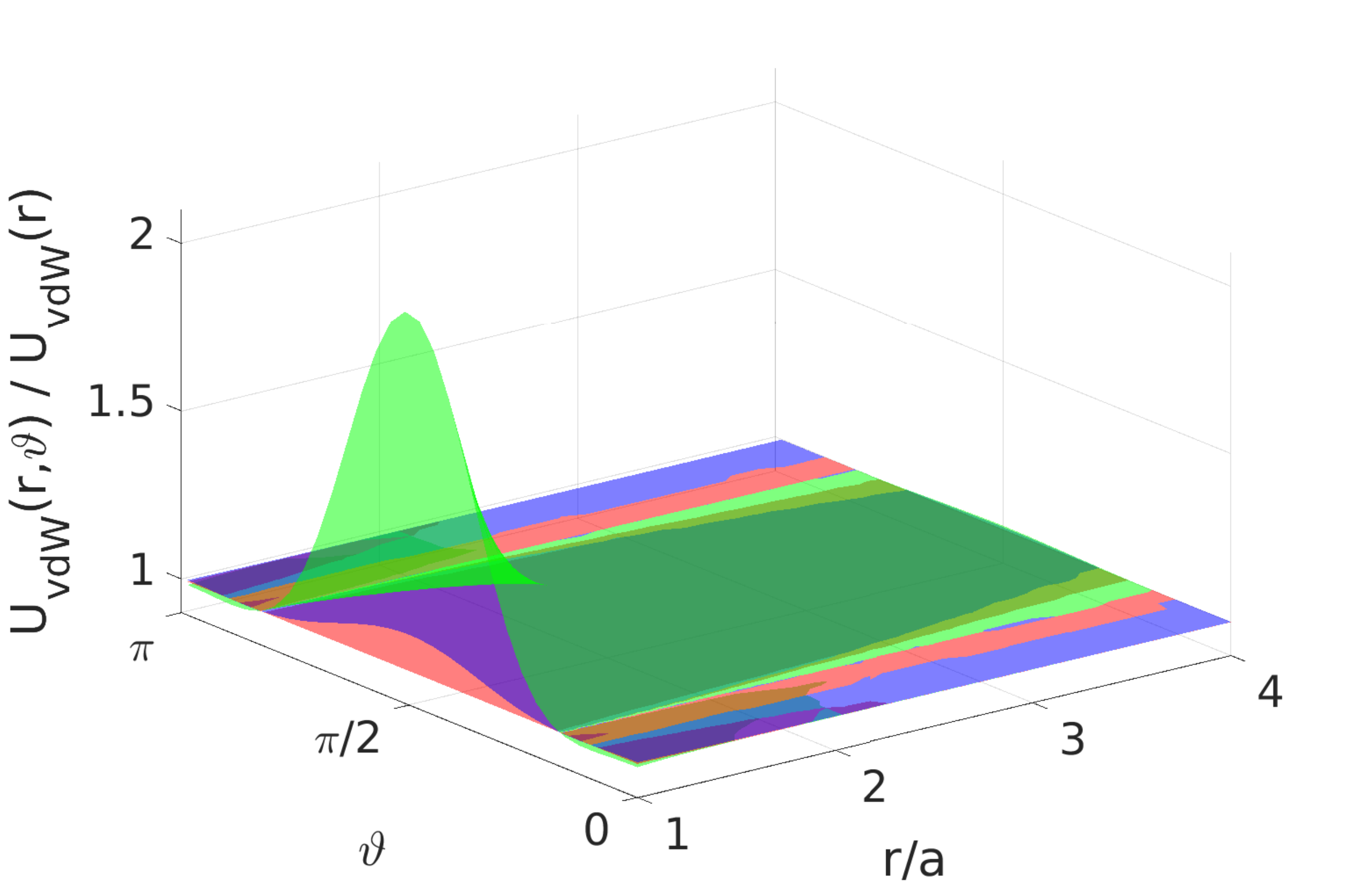}
 \caption{Relative influence of the extension and orientation dependency on the dye molecule Phthalocyanine. The zeroth order is depicted by the red surface ($f_{0,0}$), up to the first order by the blue surface ($f_{1,0}$ and $f_{0,1}$) up to the second order by the green surface ($f_{1,1}$).} \label{fig:aniso}
\end{figure}
Figure~\ref{fig:aniso} illustrates this result for Phthalocyanine. In agreement with other calculations~\cite{Fiedler15}, one finds an increase of the interaction potential on small separation distances, here up to a factor of two. It has to be noticed that higher order of corrections are physically irrelevant because one would leave the macroscopic assumptions and has to use a microscopic model for the respective distance regime. The contributions for small separations will diverge due to the overlapping polarisability densities which were assumed. 
\section{Conclusion}

The perturbative van der Waals theory developed here provides means to go beyond the theory of van der Waals potentials between point particles in a straightforward manner to account for the combined effects from spatial extension and anisotropy. Our analytical expressions will be useful in simulations and for researchers working on different aspects of intermolecular interactions.  Attempts have in the past been made to go beyond perturbative theories. A non-perturbative theory of van der Waals interaction between two finite size, and isotropic, atoms near and in close contact was derived by Bostr{\"o}m  \textit{et al.}~\cite{1402-4896-90-3-035405}.  It was found that; while non-perturbative theory is required when two electron clouds overlap; the perturbative theory works surprisingly well when two atoms are approaching each other, even when they come into close contact.  This suggests that the perturbative quantum electrodynamics theory presented here for van der Waals interaction between two anisotropic and finite size particles provides a realistic model as long as the polarisation clouds are non-overlapping.

\begin{acknowledgments}
We thank S. Zamulko for fruitful discussions and support by illustrating our theory. 
 We gratefully acknowledge support by the German Research Council (grant BU1803/3-1, S.Y.B. and J.F.) the Research Innovation Fund by the University of Freiburg (S.Y.B. and J.F.) and the Freiburg Institute for
Advanced Studies (S.Y.B.).
We acknowledge financial support from the Research Council of Norway (Projects 250346).
\end{acknowledgments}

\appendix
\section{Exact solution}\label{app1}
The correction of the van der Waals potential (\ref{eq:full}) for arbitrary orientations of both particles can be performed analytically and results in 
\begin{eqnarray}
g&=& \operatorname{tr} \left[ {\bf{R}}\left((\beta_1^A,\beta_2^A,\beta_3^A)^T\right)\cdot \operatorname{diag} (e_1^A,e_2^A,1)\right.\nonumber\\
&&\left.\cdot{\bf{R}}^T\left((\beta_1^A,\beta_2^A,\beta_3^A)^T\right)\cdot \operatorname{diag} (-2,1,1)\right.\nonumber \\
&&\left.\cdot{\bf{R}}\left((\beta_1^B,\beta_2^B,\beta_3^B)^T\right)\cdot \operatorname{diag} (e_1^B,e_2^B,1)\right.\nonumber\\
&&\left.\cdot {\bf{R}}^T\left((\beta_1^B,\beta_2^B,\beta_3^B)^T\right)\cdot \operatorname{diag} (-2,1,1)\right]\,,\\
 f_{0,0} &=& e_1^Ae_2^Ae_1^Be_2^B\,,\\
 f_{1,0}&=& \frac{512}{9}\Biggl\lbrace \left[ \left({e_1^B}^2-{e_2^B}^2\right)\cos^2\beta_3^B +{e_2^B}^2-1\right]\cos^2\beta_2^B\Biggr.\nonumber\\
 &&\Biggl.+\frac{7-{e_1^B}^2-{e_2^B}^2}{8}\Biggr\rbrace\,,\\
  f_{0,1}&=& \frac{512}{9}\Biggl\lbrace \left[ \left({e_1^A}^2-{e_2^A}^2\right)\cos^2\beta_3^A +{e_2^A}^2-1\right]\cos^2\beta_2^A\Biggr.\nonumber\\
 &&\Biggl.+\frac{7-{e_1^A}^2-{e_2^A}^2}{8}\Biggr\rbrace\,.
\end{eqnarray}

\bibliography{mqo_AnnPhys}

\end{document}